\documentclass[12pt]{iopart}
\usepackage{graphicx}
\usepackage{xcolor}
\usepackage{mathptmx}

\usepackage{iopams}
\usepackage{bm}
\usepackage{nameref,hyperref}

\renewcommand{\bm}{\mathbf m}
\newcommand{\be}{\mathbf e}
\newcommand{\bff}{\mathbf f}
\newcommand{\bg}{\mathbf g}

\newcommand{\bG}{\mathbf G}
\newcommand{\boldeta}{\boldsymbol{\eta}}

\newcommand{\bor}{\mathbf r}

\newcommand{\modm}{|\bm|}
\newcommand{\mmtrans}{\bm \bm^T}

\newcounter{inlineenum}
\renewcommand{\theinlineenum}{\roman{inlineenum}}

\renewcommand\thesubsection{\Alph{subsection}}

\begin{document}

\title[Discovering SDEs for finite flocks]{Data-driven discovery of stochastic dynamical equations of collective motion}
 \author{
     Arshed Nabeel$^1$, 
     Vivek Jadhav$^1$, 
     Danny Raj M.$^2$,
     Cl\'ement Sire$^3$,
     Guy Theraulaz$^{1,4}$,
     Ram\'on Escobedo$^4$,
     Srikanth K. Iyer$^5$ and
     Vishwesha Guttal$^1$}

 \address{$^1$ Center for Ecological Sciences, Indian Institute of Science, Bengaluru, India.}
 \address{$^2$ Department of Chemical Engineering, Indian Institute of Science, Bengaluru, India.}
  \address{$^3$ Laboratoire de Physique Théorique, CNRS, Université de Toulouse – Paul Sabatier, Toulouse, France.}
   \address{$^4$ Centre de Recherches sur la Cognition Animale, Centre de Biologie Intégrative, CNRS, Université de Toulouse – Paul Sabatier, Toulouse, France.}
 \address{$^5$ Department of Mathematics, Indian Institute of Science, Bengaluru, India.}

\vspace{10pt}
\begin{indented}
\item[]January 2022
\end{indented}

\begin{abstract}
 Coarse-grained descriptions of collective motion of flocking systems are often derived for the macroscopic or the thermodynamic limit. However, many real flocks are small sized (10 to 100 individuals), called the mesoscopic scales, where  stochasticity arising from the finite flock sizes is important. Developing mesoscopic scale equations, typically in the form of stochastic differential equations, can be challenging even for the simplest of the collective motion models. Here, we take a novel {\it data-driven equation learning} approach to construct the stochastic mesoscopic descriptions of a simple self-propelled particle (SPP) model of collective motion. In our SPP model, a focal individual can interact with $k$ randomly chosen neighbours within an interaction radius. We consider $k=1$ (called stochastic pairwise interactions), $k=2$ (stochastic ternary interactions), and $k$ equalling all available neighbours within the interaction radius (equivalent to Vicsek-like local averaging). The data-driven mesoscopic equations reveal that the stochastic pairwise interaction model produces a novel form of collective motion driven by a multiplicative noise term (hence termed, noise-induced flocking). In contrast, for higher order interactions ($k>1$), including Vicsek-like averaging interactions, yield collective motion driven primarily by the deterministic forces. We find that the relation between the parameters of the mesoscopic equations describing the dynamics and the population size are sensitive to the density  and to the interaction radius, exhibiting deviations from mean-field theoretical expectations. We provide semi-analytic arguments potentially explaining these observed deviations. In summary, our study emphasizes the importance of mesoscopic descriptions of flocking systems and demonstrates the potential of the data-driven equation discovery methods for complex systems studies. 
\end{abstract}

%
%
%
%
%

\section{Introduction}

Collective motion is a ubiquitous phenomenon in nature, observed across scales in a wide variety of systems, from microscopic organisms, insects, fish, and mammals to human crowds and even in synthetic active matter~\cite{ramaswamy2010mechanics, ramaswamy2017active, vicsek2012collective}. Collective phenomena have been a matter of investigation from the perspective of a range of disciplines beyond biology, including physics and engineering~\cite{vicsek2012collective,sumpter2010collective,camazine2020selforg, vasarhelyi2018-optimized-flocking, burkle2011-towards-uav-swarms, berlinger2021-implicit-coord-underwanter}. A central question in the field of collective motion is to understand how the simple individual behavioural rules translate to the self-organised emergent dynamics of the group~\cite{ramaswamy2010mechanics, vicsek2012collective,  deutsch2020-multi-scale-analysis-of-collective-migration, deutsch2012-collective-motion-in-bio-systems}.

To address this question, a classic and highly successful approach is that of individual-based models, where one begins with simple rules for each individual. 
These rules, for example, may include how organisms align their direction of motion with, attract towards and/or repel from their neighbours~\cite{huth1992-simulation-fish-school-movement, Aoki1982simulation, vicsek1995novel, couzin2002collective, hemelrijk2005frontaldensity, calovi2014swarming}. Further, these models incorporate errors in decision-making of organisms in the form of noise in movement. The resulting emergent properties of the groups are then studied by computer simulations. In addition, one may analytically derive the coarse-grained description of the group properties or order parameters, such as the group polarisation, which determines the degree of directional alignment of the entire flock~\cite{toner1995-long-range-order, toner1998-flocking-theory, czirok1999-kinetice-phase-transition, romanczuk2012-mean-field-theory-velocity-alignment, grossmann2012-active-brown-particles}. Unfortunately, deriving such coarse-grained models is analytically difficult except for the simplest of models ~\cite{boettiger2018-noise-to-knowledge, bertin2013-mesoscopic-theory-active-nematics, majumder2021-finite-size-effects}. Besides, many of the coarse-grained descriptions are accurate only when the number of individuals is very large, where the stochastic fluctuations average out. 

Many real organisms, on the other hand, form groups that can be relatively small to medium-sized (10 to 100 individuals), an intermediate scale which we call the \emph{mesoscopic} scale. Many experimental studies of collective motion too consider group sizes in this range~\cite{tunstrom2013collective,jhawar2020fish, calovi2018disentangling, lei2020computationalplosbio, Gautrais2012DecipheringGroups, herbert2011inferring-pnas, katz2011inferring-pnas}. At these  scales, individual-level stochasticity can have observable effects at the group level~\cite{reynolds2021intrinsic}. And the resultant, group-level stochasticity can have unusual effects on the nature of collective motion. This is best illustrated with a recently studied example of collective motion in karimeen fish ({\it Etroplus suratensis}). Here, each individual seems to copy {\it one} randomly chosen  group member~\cite{jhawar2020fish}. Under the assumption of such a simple behavioural rule, the analytical model predicts that the deterministic thermodynamic limit is a disordered phase, with no collective synchronised movement. However, when the coarse-grained descriptions also include the noise arising from finite size effects of the group sizes, the model predicts that collective order is possible when group sizes are smaller than a threshold group size~\cite{jhawar2019bookchapter}. In other words, the schooling of fish is a consequence of the noise associated with small-sized groups, and hence is termed noise-induced schooling. Intrinsic noise could also be important in evolutionary dynamics that shapes collective motion of finite flocks~\cite{joshi2017ploscomp,joshi2018evolution}. Therefore, characterising the mesoscopic description is crucial to understanding the properties of collective motion and the role of noise in finite-sized flocks~\cite{biancalani2014prl,jhawar2019bookchapter,jhawar2020fish}.

Our current understanding of the mesoscopic descriptions of collective behaviour is largely based on simple non-spatial models. For example, theoretical studies of mesoscopic models of collective behaviour~\cite{biancalani2014prl, yates2009locust, DysonPRE2015, jhawar2019bookchapter, jhawar2020fish, jhawar2020inferring} ignore space (but see~\cite{chatterjee2019three-body-interaction}) and treat animal groups as well-mixed, i.e. any individual may interact with any other group member with equal probability. Under such assumptions, using van Kampen's system size expansion, one derives Fokker-Plank and Ito's stochastic differential equations for a coarse-grained variable such as group polarisation (or degree of consensus)~\cite{gardiner1985handbook, horsthemke1989-noise-induced-transitions}. While this approach of starting with a well-mixed system may be reasonable for small group sizes, as indeed confirmed by experiments on karimeen fish~\cite{jhawar2020fish}, it is unclear how it generalises to spatially explicit models of collective motion. In the spatially explicit framework, one considers individuals as self-propelled particles which interact only within a certain local radius. Furthermore, it is well known that in flocking systems there could be density fluctuations in space as well as merge and split dynamics of groups, meaning that individuals are not always uniformly spread in space~\cite{narayan2007long,ramaswamy2010mechanics,nair2019pre,joshi2017ploscomp}. This could mean that individuals are not randomly interacting with all members of the group. Hence, whether the results of mesoscopic models with well-mixed approximation apply to spatially explicit self-propelled particle models of collective motion remains unclear.   

In this manuscript, we construct a mesoscopic description of a self-propelled particle model of collective motion. Our self-propelled particle model is based on the classic Vicsek model but with two key differences: (i) the update rules are asynchronous and (ii) we introduce a parameter $k$ fixing the number of neighbours with which a focal individual interacts  to align its direction of motion. This simpler version of the classic Vicsek model~\cite{vicsek1995novel} is inspired from studies on animal collective motion, which demonstrate that many species may not be averaging over all of their local neighbours~\cite{lemasson2009collective-selective-attention, lemasson2013motion-social-navig, calvao2014-neigh-select-plosone}. Instead, they are likely to follow only a few of their neighbours, with various studies showing that it may be as small as one or two random (or influential) neighbours~\cite{jhawar2020fish, calovi2018disentangling, lei2020computationalplosbio, Gautrais2012DecipheringGroups, herbert2011inferring-pnas, jiang2017-identifying-inf-neighbours, hinz2017-ontogeny-simple-attracfion-rules, puckett2015time-frequency-analysis, beleyur2019modeling-bats}. We then address the difficulty of analytically obtaining coarse-grained descriptions that appropriately incorporate stochasticity by using a \emph{data-driven} equation learning approach. This state-of-the-art method enables the construction of dynamical system models from the high-resolution time series of a system variable (e.g., order parameter of collective motion)~\cite{nabeel2022pydaddy, brunton2016sindy, callaham2021langevin}. The output of such an analysis is an interpretable  \emph{stochastic differential equation (SDE)}, where both deterministic and stochastic aspects of the dynamics are explicitly constructed from the data, with minimum bias of the researcher modelling the data. The main new findings of our study are:
\begin{itemize}
\item  Stochastic pairwise interaction in the {\it local neighbourhood} can maintain high group polarisation, but only in small group sizes, via intrinsic noise-induced schooling. 
\item  Higher order positive interactions (i.e., interacting with two or more neighbours, including Vicsek-like averaging) in the local neighbourhood can also drive schooling, but they can persist even at the macroscopic limit. This type of schooling is primarily explained via deterministic forcing terms, and is thus different from the noise-induced schooling driven by pairwise stochastic interactions. 
\item While the above two qualitative results are broadly consistent with the mean-field theory (MFT), the data-derived mesoscopic equations do deviate from the MFT for the following features: \begin{itemize}
    \item[{\bf A)}] MFT predicts that the deterministic  drift term is independent of the population size $N$. However, in the spatial model, the numerical coefficients of the data-derived drift function of the spatially explicit model does exhibit a dependence on $N$.
    \item[{\bf B)}] MFT predicts that the diffusion (the strength of noise) is inversely proportion to $N$. However, in the spatial model, this relationship is sensitive to the radius of the local interaction and deviates from the MFT; especially for small to intermediate group sizes. We provide semi-analytical arguments for these observed deviations. 
\end{itemize} 
\end{itemize}


\section{A brief review of non-spatial mesoscopic models}

We first briefly review the analytical models of mesoscopic description in well-mixed, or equivalently, mean-field flocking models, where the spatial structure is either not considered at all, or where spatial correlations between individuals are completely neglected. In such cases, mesoscale descriptions for small-sized flocks with simple interaction models may be derived analytically~\cite{jhawar2019bookchapter, biancalani2014prl}. In typical flocking models, individuals interact only with those within a certain metric or topological neighbourhood. However, in a  well-mixed model, a focal individual interacts with individuals from anywhere in the flock, irrespective of its distance from it. Hence, well-mixedness renders the spatially extended nature of the flocking system irrelevant. While such a well-mixed condition is far from reality, they provide a good starting point for analytical derivations of mesoscale dynamics, giving us a baseline theoretical expectation to study the impact of an actual embedding space on the dynamics of the group. 

We present the results of well-mixed models of flocks in a two-dimensional space from Jhawar et al~\cite{jhawar2020fish}. In this approach, we completely characterize the system in terms of the orientation of each individual $i$, denoted as $\be_i = (\cos \theta_i, \sin \theta_i )$, where $\theta_i$ represents the heading angle of the individual. Individuals update their orientations at each time-step based on some interaction rules, as described below. For a group of $N$ individuals, the level of \emph{order} in the group can be characterized using a \emph{polarisation order parameter}, defined as:

\begin{eqnarray}
    \bm(t) = \frac{1}{N} \sum_{i \in 1}^N \be_i(t). \label{eq:polarisation}
\end{eqnarray}
At mesoscopic scales---unlike in the thermodynamic limit---the inherent stochasticity in the dynamics becomes significant due to the finiteness in the number of individuals~\cite{durrett1994-discrete-and-spatial}. An accurate description of the system at the mesoscale should account for these stochastic effects. Therefore, we use the framework of stochastic differential equations (SDEs). Our goal is to describe the time-evolution of the order parameter $\bm$ using a stochastic differential equation of the form, interpreted in an It\^{o} sense,
\begin{eqnarray}
    \dot \bm(t) = \bff(\bm) + \sqrt{\bG(\bm)} \cdot \boldeta(t). \label{eq:sde}
\end{eqnarray}
Here, $\bff$ is a vector function called the \emph{drift} or the force, and characterises the deterministic structure of the dynamics, e.g, the existence and stability of equilibrium points in the absence of the noise term. The function $\bG$, called the \emph{diffusion}, is a symmetric matrix function, and captures the stochastic fluctuations in the dynamics. The noise term $\boldsymbol{\eta}(t) \sim \mathcal N (0, I)$ is a Gaussian white noise vector. The square root in Eq.~\ref{eq:sde} is a \emph{matrix square root}, i.e. $\sqrt \bG$ represents the symmetric matrix $\bg$ such that $\bg \bg^T = \bg^2 = \bG$. The functions $\bff$ and $\bG$ characterise the dynamics in the following way: at time $t$, $\dot \bm (t)$ is a random vector with mean $\bff(\bm(t))$ and covariance matrix $\bG(\bm(t))$---that is, $\bff$ characterises the mean behaviour of $\dot \bm$ while $\bG$ characterises the fluctuations.
When $\bG$ is a constant matrix with no dependence on $\bm$, the noise is said to be \emph{additive} or \emph{state-independent}. When $\bG$ depends on $\bm$ the noise is said to be \emph{multiplicative} or \emph{state-dependent}. 

We consider a simple class of models, where individuals can update their orientation in the following ways:
\begin{itemize}
\item \emph{Spontaneous turning:} at a rate $r_0$, an individual may spontaneously turn and choose a random direction, i.e., the new heading angle  $\theta_i$ is drawn uniformly in $[-\pi, \pi]$.
\item \emph{Stochastic pairwise interaction ($k=1$ interacting neighbour):}  at a rate $r_1$, an individual may choose a random individual from the entire group, and copy its direction.
\item \emph{Stochastic ternary interaction ($k=2$ interacting neighbours):} at a rate $r_2$, in a group of 3 individuals (picked at random from the population), the most misaligned individual takes the direction of one of the other two.
\end{itemize}
For the above class of models, analytical derivations of the mesoscale SDEs exist in the literature~\cite{jhawar2020fish, biancalani2014prl}. 

For a \textit{pairwise interaction model} with only spontaneous turns ($r_0$) and pairwise interactions ($r_1$), the mesoscale SDE takes the form~\cite{jhawar2020fish}:
\begin{eqnarray}
    \dot \bm = - r_0 \,\bm + \sqrt{\frac{r_0 + r_1(1 - \modm)^2}{N}} I \cdot \boldeta(t) \label{eq:wm-pw}.
\end{eqnarray}
The drift term of this equation, $-r_0 \,\bm$, is linear (like the force of a spring) and would alone lead to  an exponential decay of $\bm$ to ${\bf 0}$. Therefore,  in the macroscopic limit $N \to \infty$, the system is in a disordered state. However, the strength of the diffusion term becomes larger  for smaller $N$. Further, it is maximum at $\bm = {\bf 0}$, and decreases as $\modm$ increases. Consequently, the system exhibits an order, i.e., a high polarisation with $|\bm|$ approaching values close to 1, when the system size is less than a typical group size $N_c$~\cite{jhawar2020fish, biancalani2014prl,jhawar2019bookchapter}, where $N_c\sim r_1/r_0$ when $r_0\ll r_1$ (a regime that we will consider hereafter). 

For a \textit{ternary interaction model} with only spontaneous turns and stochastic ternary interactions (and no pairwise interactions), the mesoscale SDE has the form~\cite{jhawar2019bookchapter,jhawar2020fish}:
\begin{eqnarray}
    \dot \bm = - r_0 \bm + r_2 (1 - \modm)^2 \bm + \sqrt{\frac{r_0 + r_2(1 - \modm)^2}{N}} I \cdot \boldeta(t) \label{eq:wm-ter}.
\end{eqnarray}
The drift term here is cubic and has a stable manifold at $\modm = \sqrt{1 - r_0/r_2}$. The diffusion term is similar to the one present in the pairwise interaction model, and is maximum at $\bm = 0$. The ordered state in this model is largely driven by the deterministic stable equilibria. The drift term here is reminiscent of the deterministic terms typically employed in the (simpler) field theories of Vicsek-class of models~\cite{ramaswamy2010mechanics}. 

Finally, we note that the mean-field mesoscopic theory for the pairwise interaction and ternary interaction models suggests that the drift term is independent of the group size $N$, whereas the diffusion term scales inversely with $N$. Recall that these SDEs were derived under the well-mixed assumption that every individual is equally likely to interact with every other individual at all times. This assumption is strictly equivalent to the mean-field assumption, which neglects correlations between agents. In the next section, we introduce a simple spatial extension of the above interaction models, and introduce a data-driven approach to directly obtain mesoscopic SDEs from model simulations. One of our main motivations is to assess which features of the well-mixed/mean-field model survive in the presence of local interactions and, possibly, strong correlations between agents.

\section{Spatially explicit models and the data-driven equation discovery method}

\subsection{Local alignment models with asynchronous update rules for collective motion}
We develop a simple flocking model by modifying the well known Vicsek model of collective motion~\cite{vicsek1995novel}. In our model, each agent is characterised by its orientation, $\mathbf{e}_i = (\cos\theta_i, \sin\theta_i)$, position $\mathbf{x}_i$ and moves at a constant speed, $v=0.2$. Agents move within a box of length $L$ with periodic boundary conditions, and we update the positions of agents every $\Delta t$. Recent studies have emphasised the role of the probabilistic nature of animal interactions on collective motion~\cite{bode2010-stocnoise-jtb, strombom2019asynchrony, jadhav2022-randomness}. We incorporate this via asynchronous interactions among agents and choice of neighbours, as described below.

\begin{figure}
    \centering
    \includegraphics[width=\textwidth]{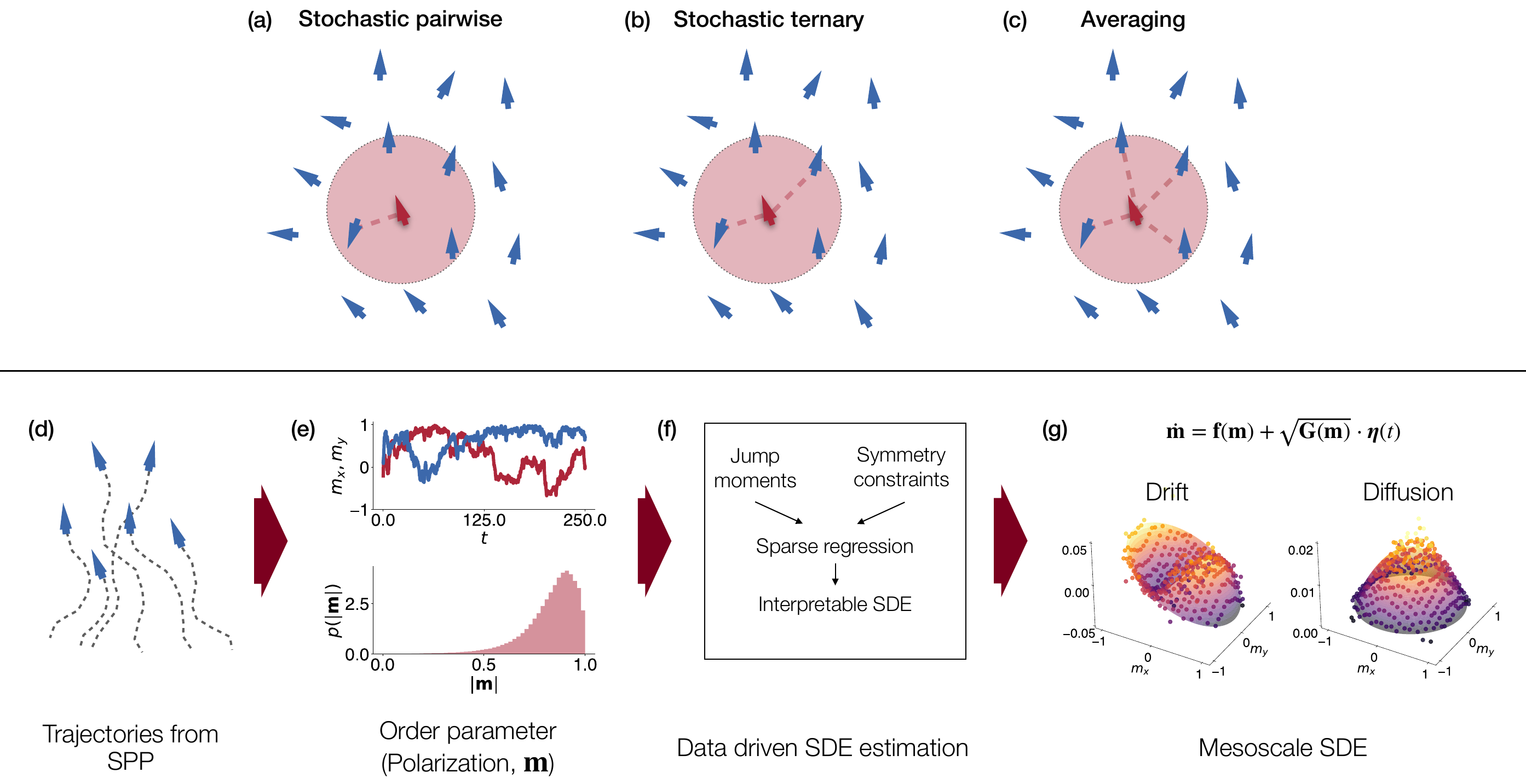}
    \caption{\textbf{Schematic illustration of the simulation model (top row: a-c) and the data-driven SDE discovery procedure (bottom row: d-g).} \textit{Top Row:} Schematic of the three local-alignment interaction models of collective motion, with asynchronous update rules. Individuals interact and align their direction of motion with others present in a circle of radius of interaction $R$, (a) with only one randomly chosen neighbour ($k=1$), (b) with two randomly chosen neighbours ($k=2$) and (c) with all neighbours in the circle ($k=$ALL). \textit{Bottom row:} (d) and (e) We simulate the model for a sufficiently long time and generate time series of the order parameter -- group polarisation $\bm$. (f) We compute jump moments and use symmetry to obtain drift and diffusion functions; we then obtain interpretable analytical functions and an SDE via sparse regression~\cite{nabeel2022pydaddy}. (g) A sample visualisation of SDE via drift and diffusion functions.}
    \label{fig:schematics}
\end{figure}

Analogous to the well-mixed models from the previous section, the agents in the spatial model also update their orientation by spontaneous turns, or by interacting with other individuals. The spontaneous turn event is identical to the one in the well-mixed model, where the individual spontaneously chooses a random direction, $\theta_i \leftarrow \eta$ where $\eta \sim \mathsf{Unif}[-\pi, \pi]$, with a rate $r_0$.
In addition, an agent may also align with its neighbour(s) within an interaction radius~$R$. We then define three models in analogy with the three well-mixed/mean-field models of the previous section (see also the top row panels of Fig.~\ref{fig:schematics}): 
\begin{enumerate}
    \item {\it Local stochastic pairwise interaction ($k=1$ interacting neighbour):} at a rate $r_1$, the focal agent copies the direction of a randomly chosen neighbour from the set of neighbours that are within the interaction radius. 
    \item {\it Local stochastic ternary interaction model ($k=2$ interacting neighbours):} at a rate $r_2$, the focal individual takes the average direction of two randomly chosen neighbours within the interaction radius. 
    \item {\it Local averaging ($k={\rm ALL}$ interacting neighbours):} at a rate $r_A$, the focal agent takes the average direction of all neighbours within the interaction radius. 
\end{enumerate}

For each of the interaction models described above, the other two interactions are absent for that model. Similar to the well-mixed models, these alignment interactions happen asynchronously and stochastically, with rates $r_0$ (spontaneous turns), $r_1$ (pairwise), $r_2$ (ternary), and $r_A$ (averaging). The reader might notice that the local averaging model is simply an asynchronous counterpart of the classic Vicsek model~\cite{vicsek1995novel}. In these models, probabilistic interaction rules are implemented asynchronously across individuals, as opposed to synchronous updates at each time-step like in the Vicsek model. We choose this asynchronous variant instead of the vanilla Vicsek model, as several previous studies have derived the underlying SDE for simple non-spatial pairwise and ternary stochastic models, like the ones presented in the previous section~\cite{biancalani2014prl,jhawar2019bookchapter,yates2009locust, DysonPRE2015}. Furthermore, asynchronous update rules are biologically more likely. Therefore, an asynchronous counterpart of the Vicsek model is more appropriate to make comparisons with the well-mixed models more direct.

The models have the following parameters: the number of agents $N$, the simulation area $L \times L$, the local radius of social interaction $R$, the spontaneous turning rate $r_0$, the pairwise alignment interaction rate $r_1$, the ternary alignment interaction rate $r_2$, and the rate $r_A$ of local-averaging among all neighbours within a radius $R$.  We choose $N$ in the range $5-80$, which covers the typical range of experimental studies.  When we vary $N$, we consider two scenarios: a constant simulation arena size, for which we choose $L=5$; and a constant density case, for which the density $\rho=N/L^2$ is fixed to be 1.2 particles per unit area.  For most of our study, we fix $R=1$. But to also study the effect of $R$, we also study simulations with $R=1.5$ and $2$ (see \emph{Results} section). The interaction rates were chosen so that, for $N=30$, the average magnitude of the polarisation across the 3 models was approximately $\modm\approx 0.8$. With these considerations, we choose $r_1=1.5$, $r_2=1$, and $r_A=1.22$, and the corresponding spontaneous turning rates for the three models were $r_0=0.014$, $0.049$, and $0.15$, respectively. We reiterate that when we consider a given interaction model, the other interaction rates are zero: for example, for the ternary interaction model, the pairwise and average copying rules are absent. 
Each simulation begins with random orientation of individuals placed roughly at the centre of the $L \times L$ continuous two-dimensional space. The simulations continue for a duration of $10^5$ time units, with asynchronous update rules ~\cite{gillespie-1976-jcp, gillespie-1977-acs}. We assume periodic boundary conditions. 



\subsection{Data-driven approach for deriving mesoscopic descriptions}
To describe the mesoscale dynamics of the different models under study, we use a data-driven approach. This general procedure consists of the following steps (see Fig.~\ref{fig:schematics})):

\begin{itemize}
    \item First, we generate simulated trajectories using the spatial models described above.
    \item Next, we quantify the dynamics of the system using an appropriate \emph{order parameter}, which characterizes the state of the system. In our case, the order parameter of interest is the group polarisation.
    \item Finally, using a data-driven procedure, we find an appropriate stochastic differential equation model to describe the dynamics of the order parameter.
\end{itemize}




From the individual trajectories of the  time series of the order parameter $\bm$, obtained from empirical observation or simulations (as in our case), one can discover an SDE model by computing the \emph{jump moments}~\cite{tabar2019book}. Briefly, given the polarisation time series $\bm(t)$, sampled at some finite sampling time $\Delta t$, the first jump moment is an estimate of the drift $\bf$:
\begin{equation}
    F(\tilde \bm) = \left \langle \bm(t + \Delta t) - \bm(t) \right \rangle_{\bm(t) = \tilde \bm}.
\end{equation}
Once $\bff$ is estimated, the diffusion can be estimated from the residuals as follows:
\begin{eqnarray}
    \bor(t) = \left( \bm(t + \Delta t) - \bm(t) \right) - \bff(\bm(t)), \\
    G(\tilde \bm) = \left \langle \left( \bor(t + \Delta t) - \bor(t) \right )\left( \bor(t + \Delta t) - \bor(t) \right )^T \right \rangle_{\bm(t) = \tilde \bm}.
\end{eqnarray}
To find interpretable expressions for $\bff$ and $\bG$, we use sparse regression to fit them as polynomial functions of $\bm$, broadly following the protocols described in \cite{nabeel2022pydaddy}. We make an important modification to take advantage of the symmetries of $\bm$. Since the individuals do not have a preferred direction in any of the models, $\bm$ must exhibit rotational and mirror symmetry---the drift and diffusion functions should respect these symmetries. Therefore, the drift function can be expressed as a function of $\bm$ and $\modm$, while the diffusion function can be expressed as a function of $\mmtrans$ and $\modm$. Thus, we express the discovered drift functions as a ``vector polynomials" with terms $\bm, \, \modm\bm, \, \modm^2\bm, \ldots$, and the diffusion functions as matrix polynomials with terms $I, \, \modm I, \, \modm^2 I, \, \ldots \, , \,  \mmtrans, \modm \mmtrans, \, \modm^2 \mmtrans, \ldots$, utilising the identity that $(\mmtrans)^{(n+1)} = \modm^{2n} \mmtrans$. If $\bG$ contains only $\modm$-terms, $\bG$ is a diagonal matrix and the diffusion is isotropic for all values of $\bm$. Non-zero off-diagonal entries in $\bG$, which in turn causes the diffusion to be anisotropic, can only appear through terms proportional to $\mmtrans$. For the models considered here, the contribution of $\mmtrans$ terms is negligible, and can be ignored. 

\section{Results}

\subsection*{Contrast between collective motion from local stochastic pairwise interactions and higher-order interactions}

\begin{figure}
    \centering
    \includegraphics[width=\textwidth]{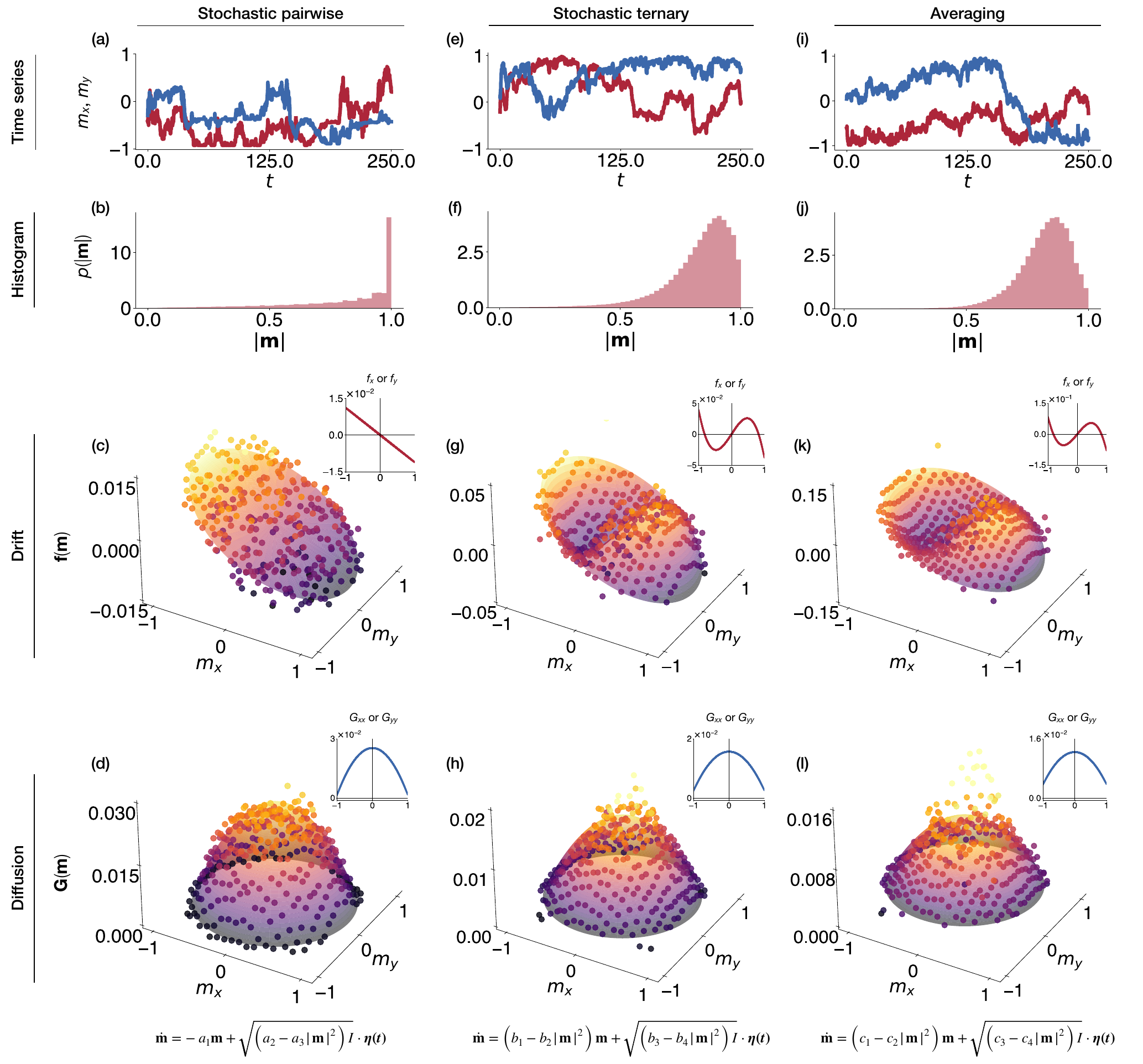}
    \caption{\textbf{Estimating mesoscopic SDEs qualitative differ between stochastic pairwise and higher-order interactions.} \textit{Top row (a, e, i)}: Sample time series of group polarisation for the three interaction models. \textit{Second row (b, f, j)}: Histograms of the net polarisation, $\modm$. We consider parameters such that all interaction models show a high degree of polarisation. \textit{Third row (c, g, k)}: The estimated drift functions via jump moments are qualitatively different between pairwise stochastic (linear) and higher-order interactions (cubic or cubic-like with three roots). The insets show a slice along the $m_x$ axis of the $x$-component $f_x$, i.e. $f_x(m_x, 0)$. \textit{Fourth row (d, h, i)}: The estimated diffusion functions are all qualitatively similar. Insets show a slice along the $x$-axis, i.e. $G_{xx}(m_x, 0)$. Bottom row equations show the estimated mesoscopic equations as interpretable SDEs. \textbf{Parameters}: $N=30$, $L=5$, $R=1$ for all three interaction models. \textbf{Estimated coefficients of SDEs}: Local stochastic pairwise: $a_1 = -0.11, a_2 = 0.023, a_3 = 0.022$. Local stochastic ternary: $b_1 = 0.081, b_2 = 0.120, b_3 = 0.016, b_4 = 0.013$. Local averaging: $c_1 = 0.171, c_2 = 0.251, c_3 = 0.012, c_4 = 0.009$.}
    \label{fig:sdes}
\end{figure}

In Fig.~\ref{fig:sdes}, we display results for three key interaction models of the spatially explicit alignment model -- stochastic pairwise, stochastic ternary and Vicsek-like local averaging. We have considered a small group size of $N=30$ to illustrate the novel features of mesoscopic dynamics. 
In Fig.~\ref{fig:sdes}, the time series of panels (a, e, i) and histogram of panels (b, f, j)  of the order parameter, i.e., the group polarisation ($\mathbf m$), show that all three models can exhibit collective motion with high directional alignment between agents. However, the underlying dynamical equations reveal interesting contrasts. 

For the stochastic pairwise interaction model ($k=1$), our data-driven discovery method yields a linear drift (c) and a quadratic diffusion~(d). The corresponding mesoscopic equation for the group polarisation is 
\begin{equation} \label{eq:pairwisemeso}
    \dot \mathbf m = -a_1 \mathbf m + \sqrt{a_2 - a_3 |\mathbf m|^2} I \cdot \boldsymbol \eta (t),
\end{equation}
where we interpret the SDE in the {\it It\^{o}-sense}, $\eta(t)$ is a standard Gaussian white noise and $a_1, a_2$ and $a_3$ are parameter related to the interaction rates. The exact values of the coefficients depend on the model parameters, the values for an exemplar case with $N = 30, R = 1$ is given in Fig.~\ref{fig:sdes}. This SDE suggests that, in the absence of noise, the system reaches the equilibrium $\mathbf m = {\bf 0}$ and thus becomes disordered. However, because of the multiplicative nature, the strength of the noise is maximum when the system is in the disordered state. Thus, the stochasticity pushes the system away from the disorder, towards the order, leading to a noise-induced high polarisation in this model. Therefore, the mesoscopic dynamics of the spatially explicit system with local pairwise interactions is qualitatively similar to the mesoscopic SDE of the corresponding well-mixed system (Eq.~\ref{eq:wm-pw}).

In contrast, we find that the mesoscopic description of the local stochastic ternary interactions ($k=2$) is of the form:
\begin{equation} \label{eq:ternarymeso}
    \dot \mathbf m = (b_1 - b_2 \mathbf{|m|}^2) \bm + \sqrt{b_3 - b_4 |\mathbf m|^2} I \cdot \boldsymbol \eta (t),
\end{equation}
where the mathematical symbols follow the same definitions as before. The drift term is a cubic function (g) whereas the diffusion term is a quadratic function (h).  Further, the collective motion is primarily driven by the drift or deterministic term, even with no or little stochasticity. Thus, the collective motion in ternary interaction systems is fundamentally different from the corresponding term of the stochastic pairwise interaction system. All these observations are also true for the mesoscopic dynamics of the well-mixed ternary interactions, whose governing equation is given by Eq.~\ref{eq:wm-ter}. In other words, the mesoscopic dynamics of the spatial system with local ternary interactions are qualitatively similar to the corresponding well-mixed system.

Finally, the mesoscopic description for the Vicsek-like local-averaging interaction model ($k={\rm ALL}$) has a qualitatively similar drift and diffusion as the stochastic ternary interaction system (see panels (k) and (l) in Fig.~\ref{fig:sdes}), with the mathematical form
\begin{equation} \label{eq:averagemeso}
    \dot \mathbf m = (c_1 - c_2 \mathbf{|m|}^2) \bm  + \sqrt{c_3 - c_4 |\mathbf m|^2} I \cdot \boldsymbol \eta (t).
\end{equation}
Technically, higher order polynomials (higher than cubic) can also give a good fit for the drift function during the sparse regression.
However, the higher-order terms do not change the number roots or the qualitative nature of the drift function in comparison to a cubic drift. Therefore, we constrain the fitting to cubic polynomials in our final fit, which gives the most parsimonious explanation for the qualitative shape and the stability structure of the drift function.



\subsection*{Diagnostics of the discovered models}

\begin{figure}
    \centering
    \includegraphics[width=\textwidth]{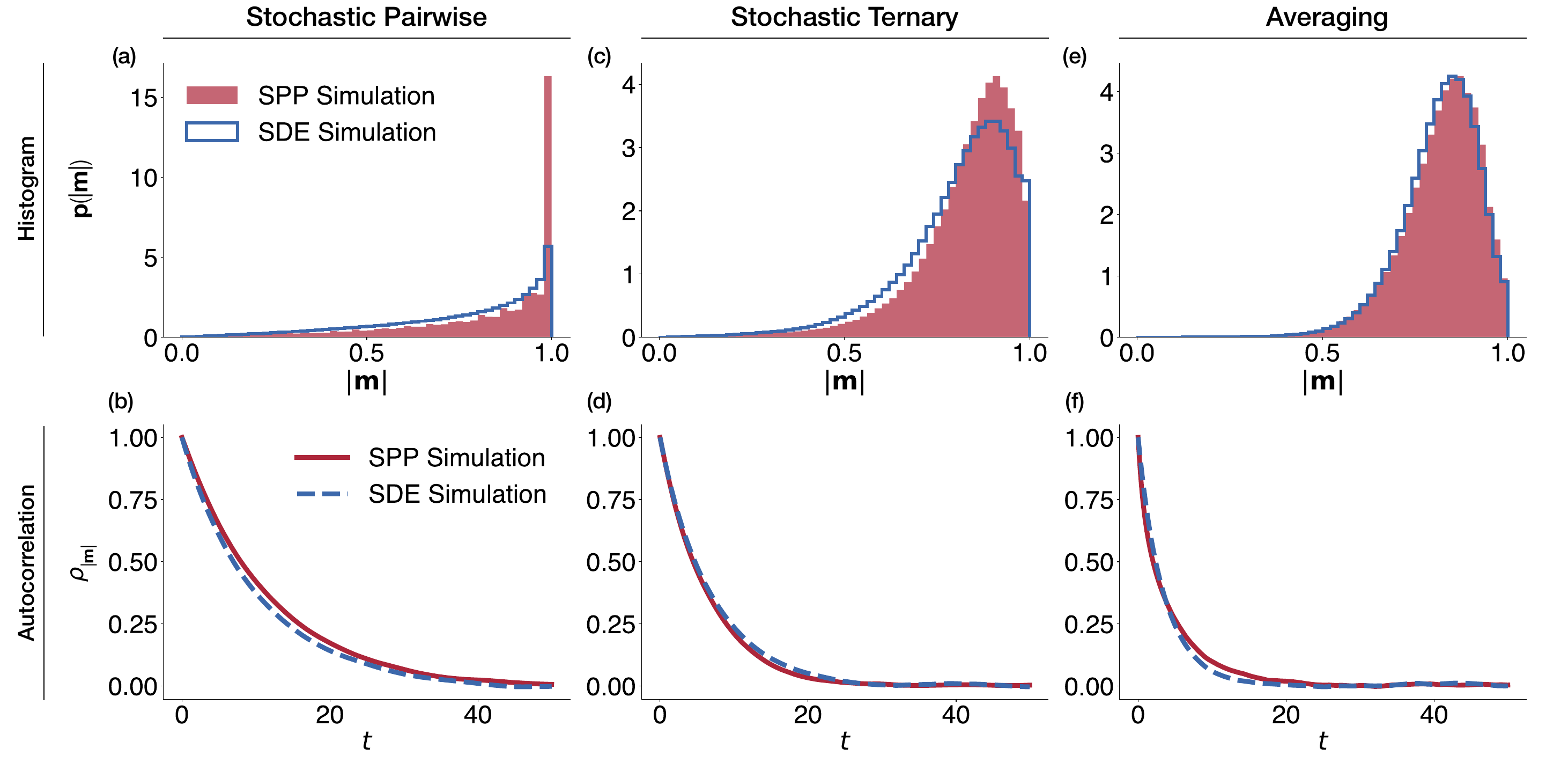}
    \caption{\textbf{Consistency of the estimated SDE models.} The data-driven mesoscale SDEs (Eqns~\ref{eq:pairwisemeso}-\ref{eq:averagemeso}) produce dynamics that closely match the actual mesoscale dynamics of the SPP models. The panels compare the distribution of the polarisation (a, c, d) and the autocorrelation functions of the polarisation (b, d, f) obtained from time series for the three SPP models and for their corresponding SDE.}
    \label{fig:consistency}
\end{figure}

We now test if the equations that we discovered via the data-driven method capture features of the data from the spatially explicit model. We simulate the discovered equations Eqs.~\ref{eq:pairwisemeso}-\ref{eq:averagemeso} using the Euler-Maruyama numerical integration scheme for Ito SDEs. In Fig.~\ref{fig:consistency} top row, we find that the histogram of the order parameter for the three spatial models and the histogram for the corresponding SDE model match reasonably well. Next, as shown in Fig.~\ref{fig:consistency} bottom row, the autocorrelation function of the order parameter also shows strong consistency between the original simulations and the SDE simulated data. Finally, we check the model consistency, as proposed by~\cite{jhawar2020inferring}. We reestimate the SDEs from the simulated SDE data. Indeed, we recover the original SDEs for each of the mesoscopic models. Therefore, we conclude that the data-driven discovery method has yielded reasonable mesoscopic SDEs for all the three spatial interaction models. 

\subsection*{Deviations of the discovered models from their well-mixed counterparts}

In the previous sections, we have observed that the discovered SDEs for both the pairwise and stochastic models are qualitatively similar to their well-mixed counterparts (compare  Eq~\ref{eq:wm-pw} to Eq~\ref{eq:pairwisemeso} and Eq~\ref{eq:wm-ter} to Eq~\ref{eq:ternarymeso}-\ref{eq:averagemeso}). However, we observe a deviation from the well-mixed results in how the parameters in the SDEs change as the number of individuals in the group increases.
Recall that the mesoscopic theory of the well-mixed systems predicts that the drift term does not depend on the $N$, while the diffusion term is inversely proportional to $N$ (see Eq.~\ref{eq:wm-pw}-\ref{eq:wm-ter}). 

We study the effect of group size on the discovered SDEs when $N$ is varied in two different ways. The first way, which is reminiscent of how real-world experiments are done, is to vary $N$ while keeping the arena size $L$ constant. This approach means that the density of particles will increase with $N$, which makes it hard to disentangle the effect of $N$ from density effects. As an alternative, we can vary $N$ and $L$ by keeping the density $\rho=N/L^2$ constant. This approach helps us to separate the effect of $N$ from the effect of density variations. We report results from both of these approaches.

\paragraph{Effect of the group size on the drift term.} 

\begin{figure}
    \centering
    \includegraphics[width=\textwidth]{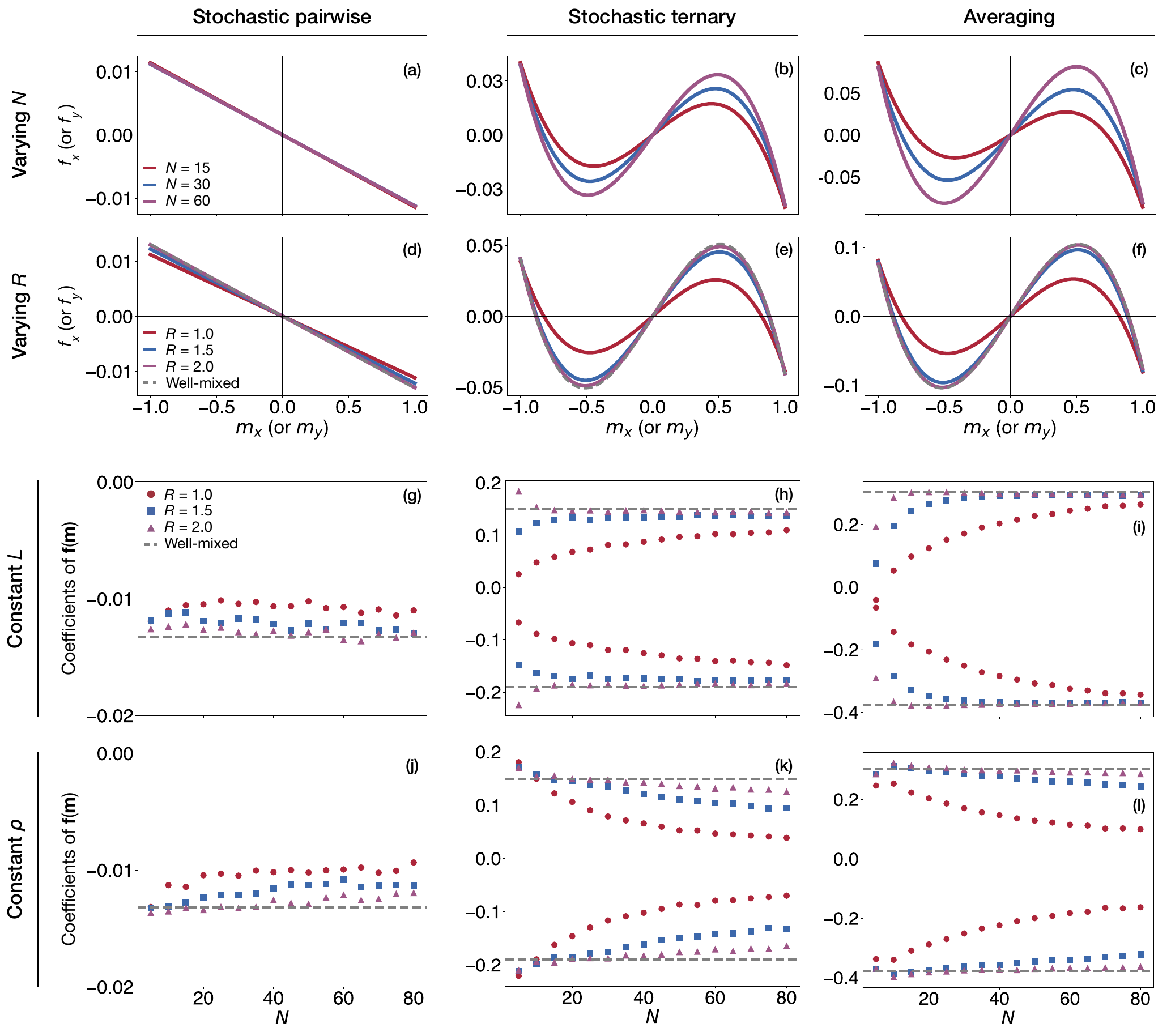}
    \caption{\textbf{Dependence of the deterministic drift  ${\bf f}\bf(\bm)$ on $N$ and $R$.} In panels (a) to (f), the drift term, $f_{x}(\bm)$, is plotted as a function of $m_x$ for the three models (same dependence for $f_{y}(\bm)$ as a function of $m_y$, by isotropy). 
    (a-c) Dependence of the drift function on $N$ for the three different models, when $R$ is constant ($R = 1$). Deviating form the mean-field theory, the drift functions change with varying $N$.
    (d-f) Dependence of the drift function on the radius of interaction, $R$. As $R$ increases, the drift functions converge to the well-mixed limit $(R = \infty)$.
    Panels (g) to (l) show the coefficients of the drift function for the 3 models (see Eqs.~\ref{eq:pairwisemeso}-\ref{eq:averagemeso}): $-a_1$ for the pairwise model in (g, j), $b_1>0$ and $-b_2<0$ in (h, k) for the ternary model, $c_1>0$ and $-c_2< 0$ in (i, l) for the local-averaging model.In (g, h, i), $N$ is increased while keeping the box size $L=5$ constant, and in (j, k, l), $N$ and $L$ are increased simultaneously  such that the density $\rho = N/L^2=1.2$ remains constant. For each condition in panels (g) to (l), the drift parameters are plotted for 3 different interaction radii, $R = 1.0$, $1.5$ and $2.0$.}
    \label{fig:drift-scaling}
\end{figure}

In the well-mixed model, the drift term is independent of the group size. However, for the spatial models, we find that the coefficients of the drift term vary as a function of the group size and the density (see  Fig.~\ref{fig:drift-scaling}). For any fixed value of $N$, the drift coefficients approach the mean-field values as the interaction radius $R$ increases. In fact, for $R \geq L/\sqrt{2}$, the model converges to the well-mixed model, and the coefficients converge to the well-mixed limit.
We speculate that this deviation is due to the fact that the effective interaction rates in the spatial models vary as a function of $N$ and $R$: this is consistent with the observation that the drift term for the pairwise interaction model---which in theory depends only on the stochastic turning rate $r_0$---stays independent of $N$ for the spatial models. Below, we propose a scaling argument in order to interpret the variation of the effective parameters of the SDE models as a function of $N$, the density $\rho$, and the interacting radius $R$.

\begin{figure}
    \centering
    \includegraphics[width=\textwidth]{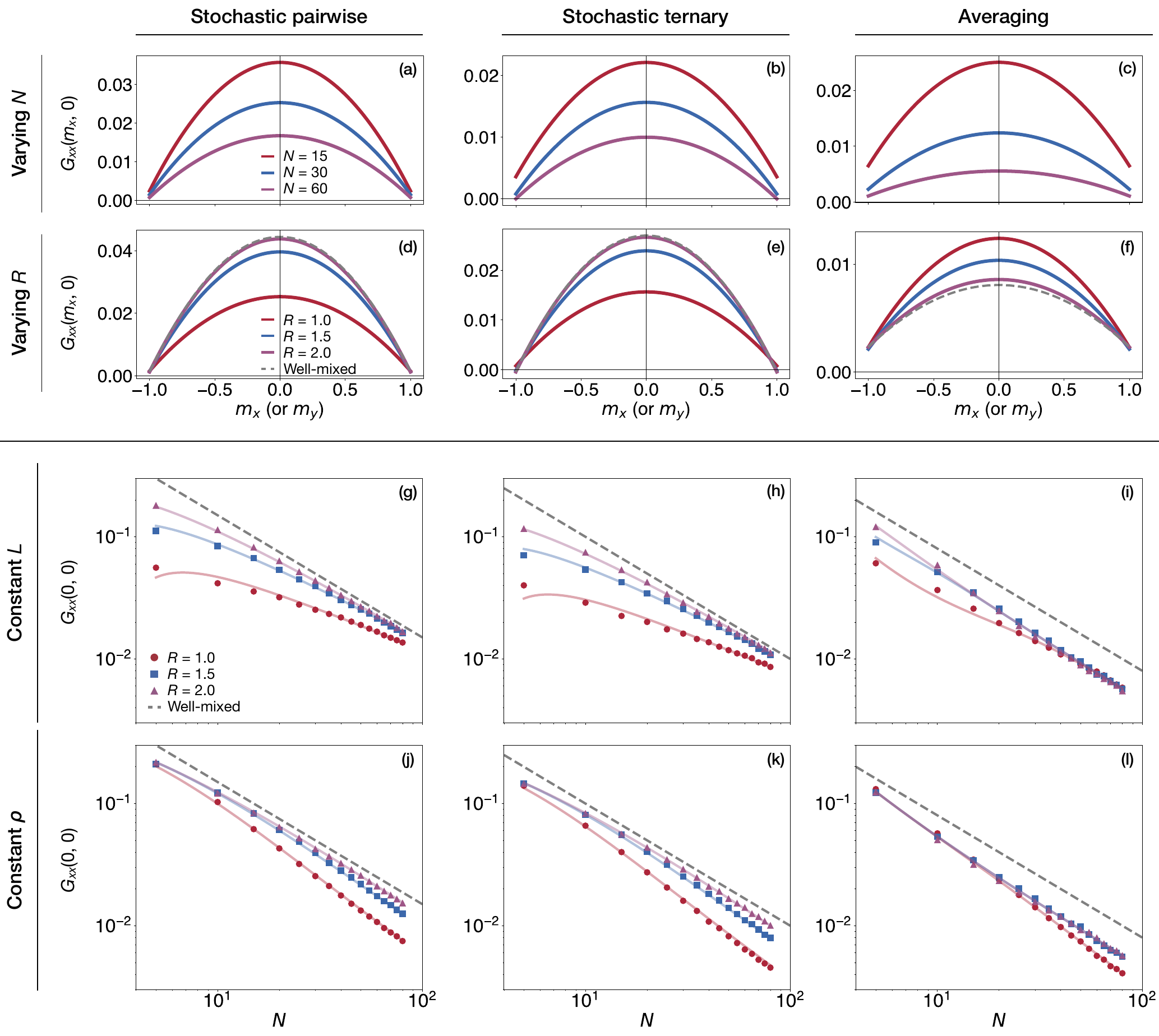}
    \caption{\textbf{Dependence of the diffusion term $\bG(\bm)$ on $N$and $R$.} 
    In panels (a) to (f), the diffusion, $G_{xx}(\bm)=G_{yy}(\bm)$, is plotted as a function of $\modm$ for the three models, presenting the inverse parabolic form of  Eqs.~\ref{eq:pairwisemeso}-\ref{eq:averagemeso}. Panels (a, b, c) and (d, e, f) illustrate the dependence of the diffusion on $N$ (for $R=1$) and on $R$ (for $N=30$), respectively.
    In panels (g) to (l), the maximum diffusion strength, $G_{xx}(0, 0)=G_{yy}(0, 0)$, is plotted as a function of $N$ for the three models. In (g, h, i), $N$ is increased while keeping the box size $L=5$ constant. In (j, k, l), $N$ and $L$ are increased simultaneously  such that the density $\rho = N/L^2=1.2$ remains constant. For each condition, $G_{xx}(0, 0)$ is plotted for 3 different interaction radii, $R = 1.0$, $1.5$ and $2.0$. The full lines correspond to the fit to the scaling ansatz of Eq.~\ref{eq:scaling}, which explains the above results in terms of a cross-over between a well-mixed and a non-mean-field regime. Overall, the values of the mean number of agents in the interaction circle of radius $R$, $N_{\rm Int}=\pi\rho R^2$, span the interval 0.6--40.}
    \label{fig:scalingG}
\end{figure}

\paragraph{Effect of the  group size on the diffusion term.}

In Fig.~\ref{fig:scalingG}, we explore the  diffusion term, and in particular its maximum value reached at ${\mathbf m}={\mathbf 0}$, $G_{xx}(0, 0)=G_{yy}(0, 0)$ (the maxima of the parabolas in the panels (a) to (f) of Fig.~\ref{fig:scalingG}; see also Eqs.~\ref{eq:pairwisemeso}-\ref{eq:averagemeso}). Again, we either increase $N$ while keeping the simulation arena size $L=5$ fixed, or increase $N$ while keeping the density $\rho=N/L^2=1.2$ fixed. 
As expected from the well-mixed/mean-field models, the strength of the diffusion decreases with increasing $N$. However, the decay of the diffusion seems to deviate from the simple $1/N$ scaling predicted by the well-mixed/mean-field models, even suggesting a possible asymptotic power-law decay $G_{xx}\sim N^{-z}$, with an exponent $z$ which could depend on $\rho$ and/or $R$. Moreover, as $R$ increases, the $1/N$ decay is ultimately recovered.

Although anomalous exponents cannot be readily excluded, it is possible to understand the complex behaviour observed in Fig.~\ref{fig:scalingG} by exploiting a scaling argument describing the cross-over between a well-mixed/mean-field regime when the interaction radius $R\gg R_c(N)$ and a regime when the mean-field results do not strictly apply, for $R\ll R_c(N)$. Here, $R_c(N)$ is a cross-over length separating these two regimes, and for a given density $\rho$, $R_c(N)$ is expected to increase with $N$. Yet, in both regimes, we will now show that our data are compatible with a diffusion term scaling like $1/N$. In fact, unless extremely long-ranged correlations are present (e.g., decaying as a small enough power-law of the distance between two agents), the law of large numbers ensures that the diffusion terms should decay like $1/N$. 

Let us consider the rescaled diffusion, $g=N\times G_{xx}(0, 0)$, which should be a function of $N$ (obviously, from Fig.~\ref{fig:scalingG}) and of the dimensionless combination $N_{\rm Int}=\pi\rho R^2$. $N_{\rm Int}$ can be simply interpreted as the expected number of agents in the interaction circle of radius $R$. We propose the scaling form
\begin{equation}
    g(\rho R^2,N)=g_{\rm MF}-(g_{\rm MF}-r_0)A(\rho R^2)B(R^2/R^2_c(N)),\label{eq:scaling}
\end{equation}
where $A$ and $B$ are 2 functions that we will strongly constrain hereafter. $g_{\rm MF}$ is the value taken by $g$ for the well-mixed case corresponding to the limit $N_{\rm Int}\to\infty$ (i.e., $R\to\infty$  or $\rho \to \infty$). For instance, in the mean-field model with pairwise interactions, we have $g_{\rm MF}=r_0+r_1$ (see Eq.~\ref{eq:wm-pw}), whereas in the mean-field model with ternary interactions, we have $g_{\rm MF}=r_0+r_2$ (see Eq.~\ref{eq:wm-ter}). First, for $N_{\rm Int}\to 0$ (i.e., $R=0$ or $\rho \to 0$), the agents are not interacting, so that $g(0,N)=r_0$. Plugging this result in Eq.~\ref{eq:scaling} imposes $A(0)\times B(0)=1$, and we can take $A(0)=B(0)=1$ in all generality. Moreover, from the above definition of the crossover length $R_c$, one should recover the mean-field result when $R\gg R_c$, and Eq.~\ref{eq:scaling} imposes $\lim_{u\to\infty} B(u)=0$. Finally, at fixed $\rho$ and $R$ and in the limit $N\to\infty$, we have $B(R^2/R_c^2(N))\to B(0)=1$, and the rescaled diffusion becomes independent of $N$ and takes the asymptotic form 
\begin{equation}
g(\rho R^2,\infty)=g_{\rm MF}-(g_{\rm MF}-r_0)A(\rho R^2).\label{eq:scalingNinf}
\end{equation}
Hence, the function $A$ encodes the dependence of the rescaled diffusion on $\rho R^2$, in the $N\to\infty$ limit. Of course, if we now take the limit $\rho R^2\to\infty$, $g(\rho R^2,\infty)$ must go to $g_{\rm MF}$,  and Eq.~\ref{eq:scaling} imposes  $\lim_{u\to\infty} A(u)=0$. 

In order to fit the results of  Fig.~\ref{fig:scalingG} by exploiting Eq.~\ref{eq:scaling} and using as few fitting parameters as possible, we assume simple forms  of the functions $A$ and $B$, compatible with the constraints that we obtained above. For the pairwise model and ternary models, we have used $A(u)=\exp(-au)$ and $B(u)=\exp(-bu)$, where $a$ and $b$ are model-dependent fitting constants. In addition, we have assumed a natural power law growth, $R_c(N)\sim N^{\alpha/2}/\sqrt{\rho}$, for the cross-over length. Interestingly, our fitting procedure resulted in the same $\alpha \approx 0.8$ for both models.
The reduced variable $R^2/R^2_c(N)$ appearing in Eq.~\ref{eq:scaling} can be rewritten as
\begin{equation}
\frac{R^2}{R^2_c(N)}=\frac{\pi\rho R^2}{\pi\rho R^2_c(N)}\sim \frac{N_{\rm Int}}{N^\alpha}.
\end{equation}
The model is effectively in the mean-field or well-mixed regime only when $R\gg R_c(N)$, i.e., $N_{\rm Int}\gg {N^\alpha}$, and the diffusion term then behaves like $G_{xx}(0, 0)=g_{\rm MF}/N$. Otherwise, for $R\ll R_c(N)$ (or equivalently, $N_{\rm Int}\ll {N^\alpha}$), we have  $G_{xx}(0, 0)=g(\rho R^2,\infty)/N$, where $g(\rho R^2,\infty)/N$ is given by Eq.~\ref{eq:scalingNinf}. Finally, for the model where the focal agent interacts with all other agents in the interacting circle, we find $\alpha \approx 0.4$, about half the value of the exponent for the binary and ternary interaction models.

The  result of our fitting procedure is presented in Fig.~\ref{fig:scalingG} and shows a fair agreement between the model simulations and the scaling ansatz of Eq.~\ref{eq:scaling}, and without too much effort in optimising the functional form of $A$ and $B$ to improve the fit (to keep as few fitting parameters as possible), which would anyway require to explore much larger values of $N$ and a wider range of $N_{\rm Int}$.
Again, the main purpose of this section was to make plausible the fact the diffusion scales like $1/N$, and that the complex behaviour of the diffusion observed in Fig.~\ref{fig:scalingG} can be interpreted by a scaling argument as a cross-over between a mean-field/well-mixed regime and a non mean-field regime.





\section{Discussion}

In this manuscript, we obtained mesoscopic (i.e. small-group sized) descriptions of a simple local-alignment-based model of collective motion. To do so, we adopted a novel data-driven equation learning approach~\cite{brunton2016sindy,nabeel2022pydaddy}. In the class of spatial models we considered, a focal individual interacts with $k$ randomly chosen neighbours within a radius $R$. Our results reveal broad consistency between the mean-field theory  and the spatially explicit models. However, a novel finding of our analysis is that the scaling relationship between the diffusion term or strength of noise $G$, and the group size $N$ for spatial models can depart substantially from the mean-field theory. In particular,  the considered range of $N$ and $N_{\rm Int}=\pi\rho R^2$ (the mean number of agents in the interaction circle of radius $R$) appears to have a strong impact on the scaling of the diffusion $G$. 

Our study offers insights on the collective motion of small to intermediate-sized animal groups, which have not been emphasized well enough in the literature. Much of the physics literature has focused on the thermodynamic or macroscopic limit ~\cite{vicsek2012collective, ramaswamy2010mechanics, toner1995-long-range-order, toner1998-flocking-theory}. In contrast, we focus on understanding mesoscale descriptions of biologically inspired variants of a classic collective motion model, with group polarisation as the order parameter of interest. The data-driven mesoscopic description of the order parameter yields stochastic differential equations, containing deterministic (called drift) and stochastic (called diffusion) terms. The analysis of these terms reveals that the nature of collective order at mesoscales arising from stochastic pairwise interactions ($k=1$ in our model) and stochastic ternary/higher-order interactions (i.e., $k \ge 2$ in the model) are fundamentally different. More specifically, we find that the stochastic pairwise interactions can lead to ordered collective motion at mesoscopic scales; this is due to intrinsic noise, i.e., noise arising from finite-sized systems. In contrast, for stochastic ternary or the higher order interaction models, including the Vicsek averaging interactions, the collective order is driven by the deterministic terms in the mesoscopic description; hence, the role of noise is secondary. These results of the spatially-explicit model with local interactions are broadly consistent with the previous mean-field theories and simulations of the collective behaviour models with no space.

Our analysis also reveals departures between the mean-field theory and spatial models when we consider how the drift and diffusion terms depend on the population size $N$. Mean field theory predicts that the drift term must be independent of $N$. For our spatial model, although the qualitative nature (i.e., functional form) of the drift is independent of $N$, we find that the quantitative features of the data-derived drift term do depend on $N$.  Mean-field theory also predicts that diffusion $G$ is inversely proportional to the population size $N$. In contrast, we find that this relationship follows an apparent power-law $G\sim N^{-z}$ for a range of $N$ and $N_{\rm Int}$, where $z$ can be  substantially smaller than $1$ when the radius of local interaction is small. We introduced  a simple scaling argument which interprets this phenomenon as a cross-over between a non mean-field regime (when $N_{\rm Int}\ll N^\alpha$) and a mean-field regime (when $N_{\rm Int}\gg N^\alpha$). Ultimately, for a given density and radius $R$, and hence $N_{\rm Int}=\pi\rho R^2$,  our analysis  indeed suggests that  $G\sim g(N_{\rm Int})/N$ scales like $1/N$ like in the mean-field models, albeit with a constant $g(N_{\rm Int})$ depending on $N_{\rm Int}$.

The above results could be discussed in light of empirical results of karimeen ({\it Etroplus suratensis})~\cite{jhawar2020fish} where authors found that $z=1$ well approximates the data, for a range of group sizes (15 to 60). Real fish are naturally extended in space, and they do not interact with all neighbours! Hence, it is interesting that these empirical data match the mean-field theoretical expectation. We speculate two possible reasons for the empirical finding: First, it is possible that the radius of interactions is already large enough for empirical data to converge to the mean-field expectations. A second possibility is that the Vicsek class of models are too simplistic for real-world applications. We add that these two possibilities are not necessarily mutually exclusive. We stress that further studies -- via simulations, theory and real-data analysis -- are needed to understand how space and the complexity of interactions among agents affect the deviations from the mean-field mesoscopic theory. 

We now ask if the stochastic pairwise copying of neighbours -- i.e., interacting with only one neighbour at a time -- over a period of time can be approximated as locally averaging. Both our mean field mesoscopic theory and data-driven mesoscopic equations clearly show that stochastic pairwise ($k=1$ in our model) and higher-order interactions ($k \ge 2$) are fundamentally different. The drift term for the stochastic pairwise is linear with disorder as the stable equilibrium; any observed collective order, therefore, is noise-induced. In the macroscopic limit ($N \to \infty$), this system admits only disorder.  On the other hand, the drift term for the higher-order interactions is cubic in which disorder ($\bm =0$) is unstable and ordered state ($|\bm| \approx 1$) forms a stable equilibria manifold. In the macroscopic limit ($N \to \infty$), this system admits order, which is typical of the Vicsek-class of collective motion models. Thus, the collective order in this model is primarily driven by deterministic forces. Hence, the governing equations and the dynamics of collective motion driven by stochastic pairwise and higher-order interactions are not equivalent either at the microscopic as well as at the group level. 

Finally, we make remarks about inferring local interactions among organisms based on the data-driven characterisation of the group-level dynamics captured as a stochastic differential equation, as suggested by~\cite{jhawar2020inferring}. This is an attractive proposition, since it is really difficult, if not impossible, to infer the local interactions that an organism follows from group-level data~\cite{nabeel2022-disentangling}, like time series for the group polarisation~\cite{jhawar2020fish,puckett2015time-frequency-analysis}. Based on the fact that drift functions are qualitatively different between stochastic pairwise and higher-order interactions, we may be able to distinguish between these two possibilities even if we only have group-level mesoscopic equations. However, our analysis suggests that ternary and local-averaging (involving multiple, time varying number of interactions) both yield qualitatively similar drift function. Hence, there are fundamental limits to what we can infer about local interactions based on mesoscopic equations alone. 

\section{Concluding remarks}
Deriving mean-field descriptions of collective systems is a non-trivial undertaking, even for highly simplified theoretical models. At mesoscopic scales, where one needs to incorporate finite-size effects and stochasticity, deriving mean-field models by hand becomes prohibitive even for relatively simple models. Therefore, we propose a data-driven approach to \emph{discover} the mesoscopic SDE models directly from simulated data. As we showed in this manuscript, even for a relatively simple class of collective motion models that accounted only for alignment interactions, we discovered some unexpected deviations from the mean-field theory. Real animal groups likely exhibit additional interactions, such as attraction and repulsion, and may have more complex interaction mechanisms. There are several models in the literature that aim to capture more realistic animal behaviour~\cite{couzin2002collective, calovi2018disentangling, wang2022-impact-individual-perception}. For such models, we argue that there is a massive potential to discover the mesoscopic equations for a variety of both toy models of collective motion as well as models of collective motion that account for detailed behaviours of the organisms. Indeed, for real-world systems, the data-derived stochastic dynamical equation is a powerful approach that may uncover the role of deterministic and stochastic forces in shaping the collective dynamics. Our approach is general enough to be applied to both real datasets and complicated models of collective motion, although care should be taken in choosing appropriate order parameter and the functional forms for the mesoscopic equations, and eventually, in interpreting the results. We hope our study inspires development of further theory, simulations as well as real data applications of these broad ideas.  

\section*{Acknowledgements}
VG and GT acknowledge the support of the Indo-French Centre for the Promotion of Advanced Research (project N°64T4-B), VG from the Science and Engineering Research Board, AN from the Ministry of Education for PhD scholarship, VJ from Prime Minister's Research Fellowship program, and DRM from Department of Science and Technology (DST) INSPIRE-Faculty award. G.T. also gratefully acknowledges the Indian Institute of Science for support via Infosys Visiting Chair Professor at the Centre for Ecological Sciences, IISc, Bengaluru. 

\section*{References}
\bibliographystyle{iopart-num}
\bibliography{references.bib}

\end{document}